\documentclass[journal]{IEEEtran}
\ifCLASSINFOpdf

\else

\fi
\usepackage[cmex10]{amsmath}
\usepackage{amssymb}
\usepackage{cite}
\usepackage{graphicx}
\usepackage{array,color}
\usepackage{multirow}
\usepackage{amsmath}
\usepackage{stfloats}
\usepackage{graphicx}
\usepackage{subfigure}
\usepackage{tabularx}
\usepackage{epsfig,epsf,color,balance,cite}
\usepackage{setspace}
\usepackage{bm}
\usepackage{textcomp}
\usepackage[linesnumbered, ruled]{algorithm2e}
\usepackage{subfigure}
\usepackage{caption}
\usepackage{graphicx}
\usepackage{setspace}
\SetKwRepeat{Do}{do}{while}%

\begin{document}

\title{Joint Blocklength and Location Optimization for URLLC-enabled UAV Relay Systems}
\author{ Cunhua Pan, Hong Ren,  Yansha Deng, Maged Elkashlan, and Arumugam Nallanathan
}

\maketitle
\vspace{-2cm}
\begin{abstract}
This letter considers the unmanned aerial vehicle (UAV)-enabled relay system to deliver command information under  ultra-reliable and low-latency communication (URLLC) requirements. We aim to jointly optimize the blocklength allocation and the UAV's location to minimize the decoding error probability subject to the latency requirement. The achievable data rate under finite blocklength regime is adopted. A novel perturbation-based iterative algorithm is proposed to solve this  problem. Simulation results show that the proposed algorithm can achieve the same performance  as  the  exhaustive search method, and significantly outperforms the existing algorithms.
\end{abstract}
\vspace{-0.4cm}

\IEEEpeerreviewmaketitle
\vspace{-0.6cm}
\section{Introduction}
\vspace{-0.1cm}
UAV-assisted communication has attracted extensive attention due to its fast deployment and favorable channel gain \cite{yongzengmaga}. UAVs can also serve as  relays to provide wireless connectivity between two devices without  direct communication links \cite{yongzengmaga,yongzeng2016,rongfeifan,shunqing2018,qwuJSAC}. Joint relay trajectory and power allocation was studied in \cite{yongzeng2016}. In \cite{rongfeifan}, UAV node placement and communication resource allocation were jointly optimized. In \cite{shunqing2018}, Zhang \emph{et al.} studied the joint trajectory and power optimization to minimize the outage probability. In \cite{qwuJSAC}, the throughput maximization problem was studied for a two-user broadcast channel, which can be regarded as a  decode-forward relay system where each hop has the same rate.

In 2017, URLLC has been regarded as  one of three pillar applications that should be supported in the 5G communications \cite{Shafijsac}. Applications requiring  URLLC services include factory automation, autonomous driving, remote surgery, etc. In URLLC, short packet transmission is normally selected to support the low-latency transmission \cite{Durisi2016}. In this case, the conventional Shannon's capacity based on the law of large numbers is no longer applicable. The achievable capacity under short packet regime was first derived  in \cite{Polyanskiy2010IT}, which is a complicated function of the system parameters.

Recently, \cite{jinli} and \cite{qqwu} considered the delay issues in UAV communications. Mean packet transmission delay minimization problem was studied in \cite{jinli} with multi-layer UAVs, while minimum-rate ratio for each user was considered in \cite{qqwu} to flexibly adjust the percentage of its delay-constrained data traffic. However, the latency requirement was not considered in \cite{jinli,qqwu}, and the  Shannon's capacity formula was adopted.

In this paper, we consider a downlink communication system in a frontline as shown in Fig.~\ref{systemodel}, where a central controller needs to send command information to a distant robot that performs certain reconnaissance missions in a military area. For concealment, shelters with thick cement/metal walls are built between the outside controller  and the military area. Hence, the channel gain between the controller and the robot is weak and negligible, and requires a UAV  to fly above the shelter to assist the transmission between the controller and the robot. We aim for jointly optimizing the location of the UAV and blocklength allocation of the UAV and the controller  to minimize the decoding error probability subject to the latency  and location  constraints. To solve this problem, we propose a novel perturbation-based iterative algorithm to alternatively  optimize the location and the blocklength. Our results show that the proposed low-complexity  algorithm achieves almost the same performance as the exhaustive search method, and performs much better  than the existing algorithms.



\vspace{-0.2cm}
\section{System Model}\label{system}
\begin{figure}
\centering
\includegraphics[width=2.4in]{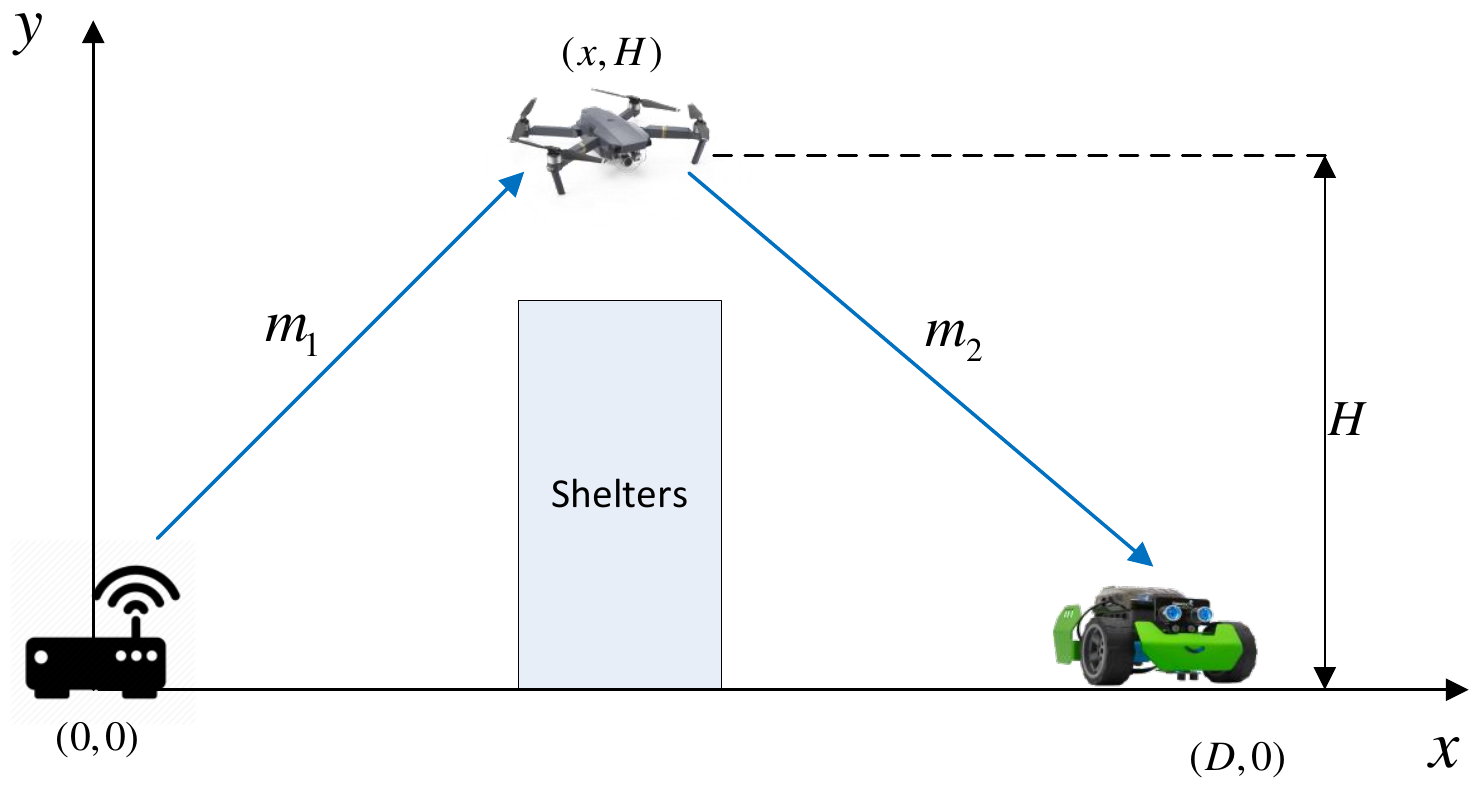}
\caption{UAV  relay system for delivering URLLC services.}
\vspace{-0.7cm}
\label{systemodel}
\end{figure}

As shown in Fig.~\ref{systemodel}, we consider a two-dimensional UAV-enabled military surveillance scenario\footnote{In this paper, we consider only  one UAV and one robot since it may be easily discovered by the enemies with more UAVs and robots. In addition, the expression of the decoding error probability for multiple UAVs and robots is very complicated and difficult to optimize, which will be left for future work.}, where the UAV hovers at $(x,H)$ above the horizontal line  between the controller and the robot\footnote{It is obvious that better system performance can be achieved with high channel power gains, which is the case when the UAV hovers above the line between the controller and the robot.}, with $H$ as the fixed altitude. The locations of  controller and  robot are $(0,0)$ and $(D,0)$. The packet size of the command signal is $L$ bits, whose transmission needs to be completed within $T_{\rm{max}}$ seconds. Then, the overall blocklength is $M=BT_{\rm{max}}$ \cite{Durisi2016}, where $B$ is the system bandwidth. Each transmission period has two phases, i.e., the first phase corresponds to the transmission from controller to UAV, while the second is from UAV to robot. The blocklength allocated for each phase is given by $m_1$ and $m_2$, respectively. The transmission powers from controller and UAV are fixed as $P_1$ and $P_2$, respectively.

The channel power gain from the controller to the UAV, and that from the UAV to the robot are denoted as ${ h_1}$ and ${h_2}$, respectively. According to the measurement result in \cite{xinlin}, the LOS probability is close to one when the UAV is above a certain altitude (e.g. 120 m), and the free space channel model can be adopted. Thus, $h_1$ and $h_2$ can be represented as
\vspace{-0.1cm}
\begin{equation}\label{edeaftg}
 {h_1} = \frac{{{\beta _0}}}{{{H^2} + {x^2}}},{h_2} = \frac{{{\beta _0}}}{{{H^2} + {{(D - x)}^2}}},\vspace{-0.1cm}
\end{equation}
where $\beta _0$ is channel power gain at a reference distance of $d_0=1$ meter.

According to \cite{Polyanskiy2010IT}, to transmit a short packet of size $L$ within $m_1$ symbols, the decoding error at the UAV is given by ${\varepsilon _1} = Q\left( {f\left( {{\gamma _1},{m_1},L} \right)} \right)$,
where $f\left( {\gamma_1, m_1, L} \right) = \ln 2\sqrt {\frac{m_1}{V_1}} \left( {{{\log }_2}(1 + \gamma_1 ) - \frac{L}{m_1}} \right)$ with $\gamma_1=P_1h_1$ \footnote{The noise is normalized to unit.} and $V_1=1-(1+\gamma_1)^{-2}$. Similarly, the decoding error probability at the robot is given by  ${\varepsilon _2} = Q\left( {f\left( {{\gamma _2},{m_2},L} \right)} \right)$, where $\gamma _2=P_2h_2$.

We consider that the UAV acts as a decode-and-forward (DF) relay. Then, the overall decoding error probability from the controller to the robot is given by
\vspace{-0.2cm}
\begin{equation}\label{dweafr}
  \varepsilon  = {\varepsilon _1} + (1 - {\varepsilon _1}){\varepsilon _2}.\vspace{-0.2cm}
\end{equation}
To enable URLLC, we aim to jointly optimize the location of the UAV and the blocklength for two phases to minimize the overall decoding error probability under the latency/blocklength constraint. Thus, the optimization problem can be formulated as
\vspace{-0.2cm}
\begin{subequations}\label{initial-pro1}
\begin{align}
\mathop {\min }\limits_{\left\{ {{m_1},{m_2},x} \right\}} \;\;\;&  {{ \varepsilon }}\\
{\rm{s.t.}}\;\;\;& d_1\le x\le d_2,\label{djreitjgot}\\
&m_1+m_2 =  M,\label{freocdsdi}\\
& m_1, m_2\in\mathbb{Z},\label{ofrrefetypo}\vspace{-0.1cm}
\end{align}
\end{subequations}
where (\ref{djreitjgot}) specifies the feasible region for $x$ that mainly depends on the shape of the shelters, $\mathbb{Z}$ is the positive integer set. Problem (\ref{initial-pro1}) is difficult to solve due to: 1) the objective function is not jointly convex w.r.t. the optimization variables; 2) even with fixed blocklength or location, the objective function is not convex w.r.t. the location or the blocklength.
\vspace{-0.3cm}
\section{Low-complexity Algorithm}\label{proformu}
In this section, we develop a low-complexity iterative algorithm to solve Problem (\ref{initial-pro1}). In particular, we first find the optimized blocklength with fixed location $x$, and then find the optimal location  by fixing the blocklength allocation, and at last alternatively solve each subproblem until convergence.
\vspace{-0.4cm}
\subsection{Optimize Blocklength Allocation with Fixed $x$}
With fixed UAV location $x$, the channel gains $h_1$ and $h_2$ are fixed according to (\ref{edeaftg}). Then, we only need to optimize the blocklength. Remind that the expression of objective function in (\ref{dweafr}) is very complicated. To handle this difficulty, we note that to guarantee the extremely low error probability for the whole link, the error probability for each link should be sufficiently small. In this case, we can approximate the overall error probability as $\varepsilon (m_1) \approx {\varepsilon _1}(m_1) + {\varepsilon _2}(m_1) \triangleq \tilde \varepsilon(m_1) $. In the following theorem, we prove that $\tilde \varepsilon(m_1) $ is a convex function of $m_1$.

\emph{\textbf{Theorem 1}}: We assume that $m_1$ is a continuous variable. Given the channel gains $h_1$ and $h_2$, $\tilde \varepsilon(m_1) $ is a convex function w.r.t. $m_1$.

\emph{\textbf{Proof}}: \upshape We first prove that  ${\varepsilon _1}(m_1)$ is a convex function w.r.t. $m_1$, then the convexity of function ${\varepsilon _2}(m_1)$ can be proved in a similar method by replacing $m_2$ with $M-m_1$. Thus, the summation of two convex functions  $\tilde \varepsilon (m_1)$ is a convex function.

To simplify the notation, we denote $f_1(m_1)={f\left( {{\gamma _1},{m_1},L} \right)}$. The first order and second order of ${\varepsilon _1}(m_1)$ w.r.t. $m_1$ are given by
\begin{eqnarray}
  \varepsilon_1'(m_1) \!\!\!\!\!\!&=&\!\!\!\!\!  - \frac{1}{{\sqrt {2\pi } }}{e^{ - \frac{{f_1^2(m_1)}}{2}}}{f_1'(m_1)},\label{djrefih}\\
\varepsilon_1''(m_1)\!\!\!\!\!\! &=&\!\!\!\!\!\! \frac{1}{{\sqrt {2\pi } }}{e^{ - \frac{{f_1^2(m_1)}}{2}}}\!\left( {{f_1(m_1)}{{\left( {{f_1'(m_1)}} \right)}^2} \!\!- \!\!f_1''(m_1)} \right).\label{dhrefi}
\end{eqnarray}
Since the first term in the brackets of (\ref{dhrefi}) is positive, we only need to prove ${f_1''}(m_1) \le 0$ so that ${\varepsilon _1}(m_1)$ is a convex function.

To this end, we first simplify the function $f_1$ as
\begin{equation}\label{fref}
 {f_1}(m_1) = {A_1}\sqrt {{m_1}} \left( {{C_1} - \frac{L}{{{m_1}}}} \right),
\end{equation}
where $A_1$ and $C_1$ are constants given by ${A_1} = {{\ln 2} \mathord{\left/
 {\vphantom {{\ln 2} {\sqrt {{V_1}} }}} \right.
 \kern-\nulldelimiterspace} {\sqrt {{V_1}} }}$ and ${C_1} = {\log _2}(1 + {\gamma _1})$, respectively.

The first and second derivative of ${f_1}(m_1)$ w.r.t. $m_1$ are
\begin{eqnarray}
 {f_1'}(m_1)&=& \frac{1}{2}{A_1}{C_1}m_1^{ - \frac{1}{2}} + \frac{1}{2}{A_1}Lm_1^{ - \frac{3}{2}},\label{djefoir}\\
  {f_1''}(m_1) &=&  - \frac{1}{4}{A_1}{C_1}m_1^{ - \frac{3}{2}} - \frac{3}{4}{A_1}Lm_1^{ - \frac{5}{2}} < 0,\label{feoigjgpj}
\end{eqnarray}
which show that ${f_1}(m_1)$ is a strictly concave function of $m_1$. Then ${\varepsilon _1}(m_1)$ is a convex function w.r.t. $m_1$. By using the similar derivations, we show that ${\varepsilon _2}(m_1)$ is also a convex function w.r.t. $m_1$. This completes the proof. \hfill\rule{2.7mm}{2.7mm}

Based on Theorem 1, we apply the bisection search method to find the optimal blocklength allocation through solving the following equation:
\begin{equation}\label{dweftrgt}
\tilde \varepsilon'(m_1) = \varepsilon _1'(m_1) + \varepsilon _2'(m_1) = 0.
\end{equation}
The details of searching for the optimal blocklength allocation are provided in Algorithm \ref{hgirjoj}.
\vspace{-0.4cm}
\begin{algorithm}
\setstretch{0.5}
  \caption{Find the Optimal $m_1$ and $m_2$ with Fixed $x$  }\label{hgirjoj}
   \textbf{Initialize}  $m_1^{\rm{lb}}=1, m_1^{\rm{ub}}=M-1$ and the error tolerance $\delta  = 0.5$;

  \While { $m_1^{\rm{ub}}-m_1^{\rm{lb}}>\delta$ }
  {Set $m_1^{\rm{mid}}=(m_1^{\rm{lb}}+m_1^{\rm{ub}})/2$.

  \eIf{${\left. {\tilde \varepsilon'(m_1)} \right|_{{m_1} = m_1^{{\rm{mid}}}}} > 0$}
  {Set $m_1^{\rm{ub}}=m_1^{{\rm{mid}}}$.}
  {Set $m_1^{\rm{lb}}=m_1^{{\rm{mid}}}$.
  }
  }
  Return  the optimal $m_1$ as $m_1^* = \mathop {\arg \min }\limits_{\left\{ {\left\lfloor {m_1^{{\rm{mid}}}} \right\rfloor ,\left\lceil {m_1^{{\rm{mid}}}} \right\rceil } \right\}} \tilde \varepsilon $ and the optimal $m_2$ as $m_2^*=M-m_1^*$.
\end{algorithm}

\vspace{-0.8cm}\subsection{Optimal Location Optimization with Fixed $m_1$ and $m_2$}
In this subsection, we aim for optimizing location $x$ with given $m_1$ and $m_2$. However, the objective function $\tilde \varepsilon(x)$ is not a convex function w.r.t. $x$. In the following, we numerically show that $\tilde \varepsilon(x)$ has only one local minimum point. Hence, there only exists only one solution that minimizes $\tilde \varepsilon(x)$.

To provide clear explanations, we define a new function $g(x) \buildrel \Delta \over = {\ln}(\tilde \varepsilon(x)) $, which has the same monotonic property of $\tilde \varepsilon(x)$. We first obtain the first and second derivative of $g(x)$ w.r.t. $x$ as follows. We define $f_1(x)={f\left( {{\gamma _1}(x),{m_1},L} \right)}$ and $f_2(x)={f\left( {{\gamma _2}(x),{m_2},L} \right)}$, where ${\gamma _1}(x)$ and ${\gamma _2}(x)$ are given by  ${\gamma _1}(x)=\frac{{{P_1\beta _0}}}{{{H^2} + {x^2}}}$ and ${\gamma _2}(x)=\frac{{{P_2\beta _0}}}{{{H^2} + {(D-x)^2}}}$, respectively.

The first derivative of $g(x)$ w.r.t. $x$ is given by
\begin{equation}\label{dewdehfi}
  g'(x) = \frac{{{{\varepsilon '_1}}(x) + {{\varepsilon '_2}}(x)}}{{{\varepsilon _1(x)} + {\varepsilon _2(x)}}},
\end{equation}
where ${{\varepsilon '_i}}(x)$ is given by
\begin{equation}\label{frefre}
{{\varepsilon '_i}}(x) = \frac{{\partial {\varepsilon _i}({\gamma _i})}}{{\partial {\gamma _i}}}{{\gamma '_i}}(x),i = 1,2,
\end{equation}
with ${{\gamma '_i}}(x)$ given by
\begin{equation}\label{dewfref}
  {{\gamma '_1}}(x) =  - \frac{{2{P_1}{\beta _0}x}}{{{{\left( {{H^2} + {x^2}} \right)}^2}}},{{\gamma '_2}}(x) = \frac{{2{P_2}{\beta _0}(D - x)}}{{{{\left( {{H^2} + {{(D - x)}^2}} \right)}^2}}},
\end{equation}
and $\frac{{\partial {\varepsilon _i}({\gamma _i})}}{{\partial {\gamma _i}}}$  given by
\begin{equation}\label{daef}
\frac{{\partial {\varepsilon _i}({\gamma _i})}}{{\partial {\gamma _i}}} =  - \frac{1}{{\sqrt {2\pi } }}{e^{ - \frac{{f_i^2({\gamma _i})}}{2}}}\frac{{\partial {f_i}({\gamma _i})}}{{\partial {\gamma _i}}}.
\end{equation}
In (\ref{daef}), $\frac{{\partial {f_i}({\gamma _i})}}{{\partial {\gamma _i}}}$ is
\begin{equation}\label{deafrg}
\vspace{-0.3cm}
  \frac{{\partial {f_i}({\gamma _i})}}{{\partial {\gamma _i}}} = \sqrt {{m_i}} \frac{{1 - \ln 2\frac{{{{\log }_2}(1 + {\gamma _i}) - \frac{L}{{{m_i}}}}}{{{{(1 + {\gamma _i})}^2} - 1}}}}{{\sqrt {{{(1 + {\gamma _i})}^2} - 1} }}.
\end{equation}

The second derivative of $g(x)$ w.r.t. $x$ can be calculated as
\begin{equation*}\label{dewfrf}
  g''(x) \!=\! \frac{{\left( {{{\varepsilon ''_1}}(x) + {{\varepsilon ''_2}}(x)} \right)\left( {{\varepsilon _1}(x) + {\varepsilon _2}(x)} \right) - {{\left( {{{\varepsilon '_1}}(x) + {{\varepsilon '_2}}(x)} \!\right)}^2}}}{{{{\left( {{\varepsilon _1}(x) + {\varepsilon _2}(x)} \right)}^2}}},
\end{equation*}
where ${{{\varepsilon ''_i}}(x)},i=1,2$ are given by
\begin{equation}\label{dweef}
  {{\varepsilon ''_i}}(x) = \frac{{{\partial ^2}{\varepsilon _i}({\gamma _i})}}{{\partial \gamma _i^2}}{\left( {{{\gamma '_i}}(x)} \right)^2} + \frac{{\partial {\varepsilon _i}({\gamma _i})}}{{\partial {\gamma _i}}}{{\gamma ''_i}}(x),
\end{equation}
with  ${{\gamma ''_i}}(x)$ given by
\begin{eqnarray*}
{{\gamma ''_1}}(x)\!\!\!\!\!&=&\!\!\!\!\! \frac{{6{P_1}{\beta _0}{x^4} + 4{P_1}{\beta _0}{H^2}{x^2} - 2{P_1}{\beta _0}{H^4}}}{{{{\left( {{H^2} + {x^2}} \right)}^4}}},\label{fjreijg}\\
 {{\gamma ''_2}}(x) \!\!\!\!\!&=&\!\!\!\!\! \frac{{6{P_2}{\beta _0}{{(D \!-\! x)}^4} \!\!+\! 4{P_2}{\beta _0}{H^2}{{(D \!- \! x)}^2}\!\! -\!\! 2{P_2}{\beta _0}{H^4}}}{{{{\left(\! {{H^2} + {{(D \!-\! x)}^2}} \!\right)}^4}}},\label{jortg}
\end{eqnarray*}
and $\frac{{{\partial ^2}{\varepsilon _i}({\gamma _i})}}{{\partial \gamma _i^2}}$ given by
\begin{equation}\label{wdefr}
 \!\! \frac{{{\partial ^2}{\varepsilon _i}({\gamma _i})}}{{\partial \gamma _i^2}}\!\! =\!\! \frac{1}{{\sqrt {2\pi } }}{e^{ - \frac{{f_i^2({\gamma _i})}}{2}}}\!\!\left(\!\! {{f_i}({\gamma _i})\!{{\left(\! {\frac{{\partial {f_i}({\gamma _i})}}{{\partial {\gamma _i}}}} \!\right)}^2} \!\!\!-\! \frac{{{\partial ^2}{f_i}({\gamma _i})}}{{\partial \gamma _i^2}}}\!\! \right).
\end{equation}

In (\ref{wdefr}), ${\frac{{{\partial ^2}{f_i}({\gamma _i})}}{{\partial \gamma _i^2}}}$ is given by
\begin{eqnarray}
  \frac{{{\partial ^2}{f_i}({\gamma _i})}}{{\partial \gamma _i^2}} \!\!&=&\!\!\!\! c_i\left( { - \frac{1}{{1 + {\gamma _i}}} - (1 + {\gamma _i})} \right)\left( {{{\left( {1 + {\gamma _i}} \right)}^2} - 1} \right) \!+ \nonumber\\
   && 3c_i(1 + {\gamma _i})\left( {{{\log }_2}(1 + {\gamma _i})\! -\! \frac{L}{{{m_i}}}} \right)\ln 2,
\end{eqnarray}
where $c_i= {{\sqrt {{m_i}} } \mathord{\left/
 {\vphantom {{\sqrt {{m_i}} } {{{\left( {{{\left( {1 + {\gamma _i}} \right)}^2} - 1} \right)}^{\frac{5}{2}}}}}} \right.
 \kern-\nulldelimiterspace} {{{\left( {{{\left( {1 + {\gamma _i}} \right)}^2} - 1} \right)}^{5/2}}}}$.

\begin{figure}
\centering
\includegraphics[width=2.4in]{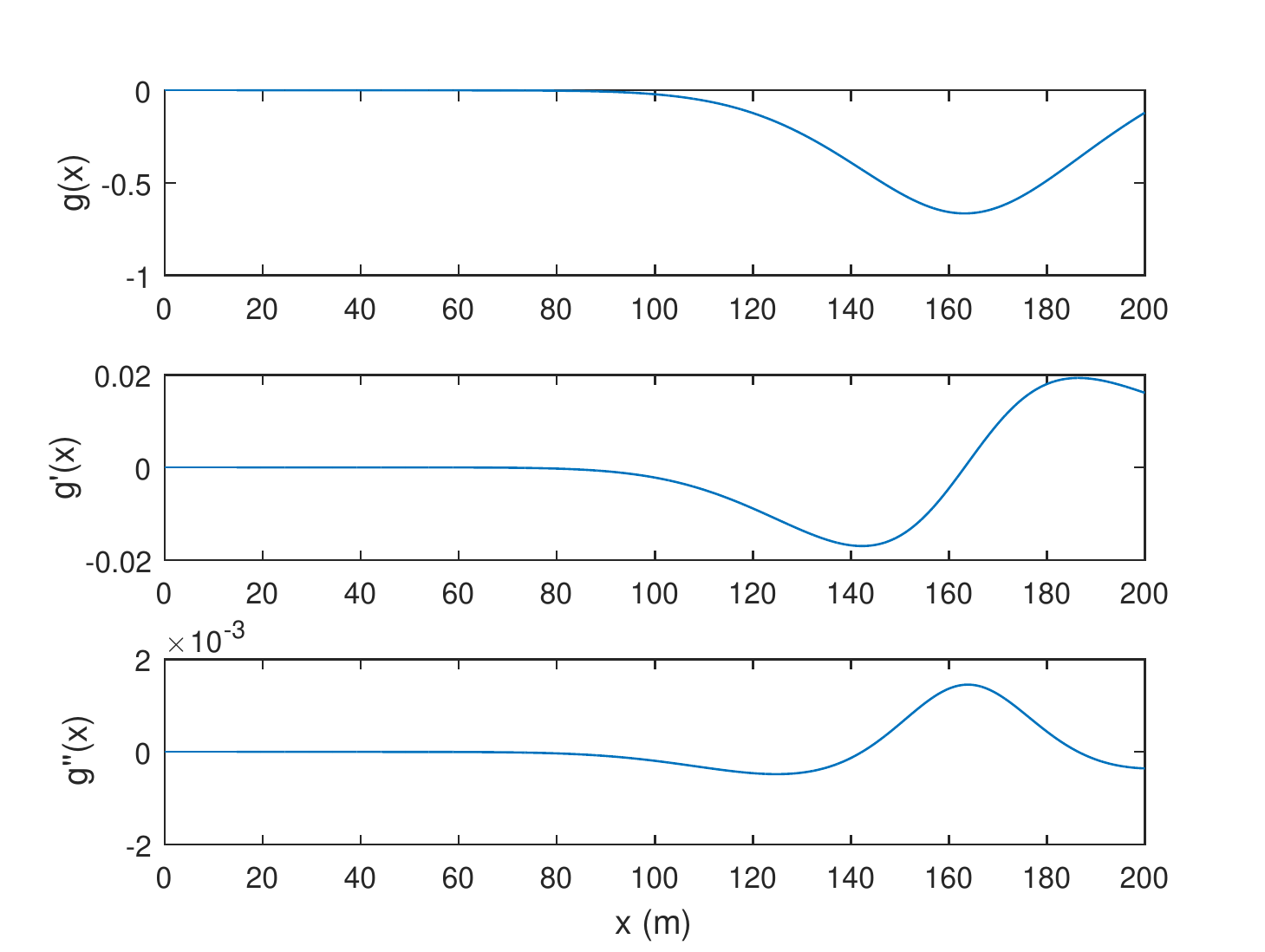}
\vspace{-0.2cm}
\caption{$g(x)$, $g'(x)$ and $g''(x)$ versus $x$, where the system parameters are the same as those in the simulation section.}
\vspace{-0.6cm}
\label{numberical}
\end{figure}

Unfortunately, $g(x)$ is not a convex function w.r.t. $x$ in the whole region of $x$. Fig.~\ref{numberical} plots the functions $g(x)$, $g'(x)$ and $g''(x)$ versus $x$. It is observed from this figure that when $0\rm{m}<x< 142.5\rm{m}$, $g''(x)<0$, which means $g(x)$ is a concave function. On the other hand, when $142.5\rm{m}<x<186.5\rm{m}$, $g''(x)>0$ and $g(x)$ is a convex function. Finally, when $186.5\rm{m}<x<200\rm{m}$, $g(x)$ becomes a concave function again. However, it is observed that $g'(x)>0$ when $0\rm{m}<x<163.4\rm{m}$, and $g'(x)<0$ when $163.4\rm{m}<x<200\rm{m}$. This means function $g(x)$ first decreases with $x$ for $0\rm{m}<x<163.4\rm{m}$ and then increases with $x$ for $163.4\rm{m}<x<200\rm{m}$, and there exists only one minimum value. It is difficult to rigorously prove this due to the cumbersome expression of $g'(x)$.  We check all the other simulation parameters, all numerical results show the same trend of function $g(x)$ \footnote{The rational behind this is that when x is small, the link from UAV to the robot will be the bottleneck of the whole link, and vice versa. }. As a result, the bisection search can be used to find the root of $g'(x)$, details of which can be found in the following algorithm.
\vspace{-0.4cm}
\begin{algorithm}
\setstretch{0.5}
  \caption{Find the Optimal $x$ with Fixed $m_1$ and $m_2$ }\label{dearfre}
   \textbf{Initialize}  $x^{\rm{lb}}=d_1, x^{\rm{ub}}=d_2$ and error tolerance $\zeta= 0.1$;

   \eIf{${\left. {g'(x)} \right|_{x = {d_2}}} < 0$}
    {Return the optimal $x^*=d_2$.}
    {\eIf{${\left. {g'(x)} \right|_{x = {d_1}}} > 0$}
         {Return the optimal $x^*=d_1$.}
         {\While{$x^{\rm{ub}}-x^{\rm{lb}}>\zeta $}
                {Set $x^{\rm{mid}}=(x^{\rm{ub}}+x^{\rm{lb}})/2$.

                \eIf{${\left. {g'(x)} \right|_{x = x^{\rm{mid}}}} > 0$}
                    {Set $x^{\rm{up}}=x^{\rm{mid}}$.}
                    {Set $x^{\rm{lb}}=x^{\rm{mid}}$.}
                }
         }
    }
\end{algorithm}

\vspace{-0.8cm}\subsection{Overall Algorithm and Analysis}\label{irehgt}
\vspace{-0.1cm}
\begin{algorithm}
\setstretch{0.5}
  \caption{Iterative Algorithm for Solving Problem (\ref{initial-pro1})}\label{caffrgf}
   \textbf{Initialize}  $m_1^{(0)}, m_2^{(0)}$ and $x^{(0)}$, integer parameter $N_{\rm{max}}$, iterative index $t=1$, maximum iterative times $t_{\rm{max}}$;

  \Repeat { $t\geq t_{\rm{max}}$ }
  {With given $x^{(t-1)}$, obtain the optimal $m_1^{(t)}$ by using Algorithm \ref{hgirjoj};\\
   Generate two random integer values $n^l$ and $n^r$ within $[1,N_{\rm{max}}]$. Use  Algorithm \ref{dearfre} to calculate the optimal location when $m_1=m_1^{(t)}-n^l, m_1^{(t)},$ $m_1^{(t)}+n^r$. The corresponding optimal $x$ is denoted as ${\cal X}^{(t)}=\{x_l^{(t)},$ $x_m^{(t)},x_r^{(t)}\}$. Choose the optimal $x$ from ${\cal X}^{(t)}$ with the minimum $\tilde \varepsilon$, and denote it as $x^{(t)}$.\\
   Set $t=t+1$;}
\end{algorithm}
It is noted from simulations that the conventional block coordinate descent method, which directly iterates between blocklength and location,  is very likely to get stuck at the initial point. To overcome this issue, we introduce a small perturbation for the blocklength $m_1$ in each iteration as shown in line 4 of Algorithm \ref{caffrgf}, where $N_{\rm{max}}$ is a small integer.

The overall algorithm for solving Problem (\ref{initial-pro1}) is provided in Algorithm \ref{caffrgf}. The convergence of this algorithm is guaranteed since the objective value decreases in each step  and the value is lower-bounded by zero. The complexity of this algorithm is analyzed as follows. In each iteration of Algorithm \ref{caffrgf}, we need to run both Algorithm \ref{hgirjoj} only once  and Algorithm \ref{dearfre} for three times. The complexity of Algorithm \ref{hgirjoj} is given by $Q_1={\cal O}\left( {{{\log }_2}(M/\delta )} \right)$, while that of Algorithm \ref{dearfre}  is $Q_2={\cal O}\left( {{{\log }_2}\left( {{{({d_2} - {d_1})} \mathord{\left/
 {\vphantom {{({d_2} - {d_1})} \zeta }} \right.
 \kern-\nulldelimiterspace} \zeta }} \right)} \right)$. Hence, the overall complexity of Algorithm \ref{caffrgf} is given by ${\cal O}\left( {{n_{\max }}({Q_1}+{Q_2})} \right)$. In simulations, the algorithm generally converges within ten iterations.

By using the similar analysis, the complexity of exhaustive search method is given by ${\cal O}\left( M(d_2-d_1)/\zeta \right)$, which is significantly higher that of Algorithm \ref{caffrgf}.

Since the original problem in (\ref{initial-pro1}) is non-convex, Algorithm  \ref{caffrgf} cannot be guaranteed to  yield a globally optimal solution. However, from the simulation results, we can find that this algorithm can achieve the same performance as the exhaustive search method.

\vspace{-0.2cm}
\section{Simulation Results}\label{simlresult}
We now perform simulation results to show the performance of our proposed algorithm. The system parameters are set as follows: system bandwidth of $B=1\ {\rm{MHz}}$, $D=200$ m, $H=120$ m, $d_1=30$ m, $d_2=130$ m, $L=100$ bits,  $M=100$, $\beta_0=50$ dB, $P_1=3$ Watt, $P_2=1$ Watt, $N_{\rm{max}}=3$, $t_{\rm{max}}=10$. The system transmission delay duration  is set as $T_{\rm{max}}=100\ {\rm{us}}$. Thus, the total number of symbols is $M=BT_{\rm{max}}=100$.

\begin{figure}
\centering
\includegraphics[width=2.4in]{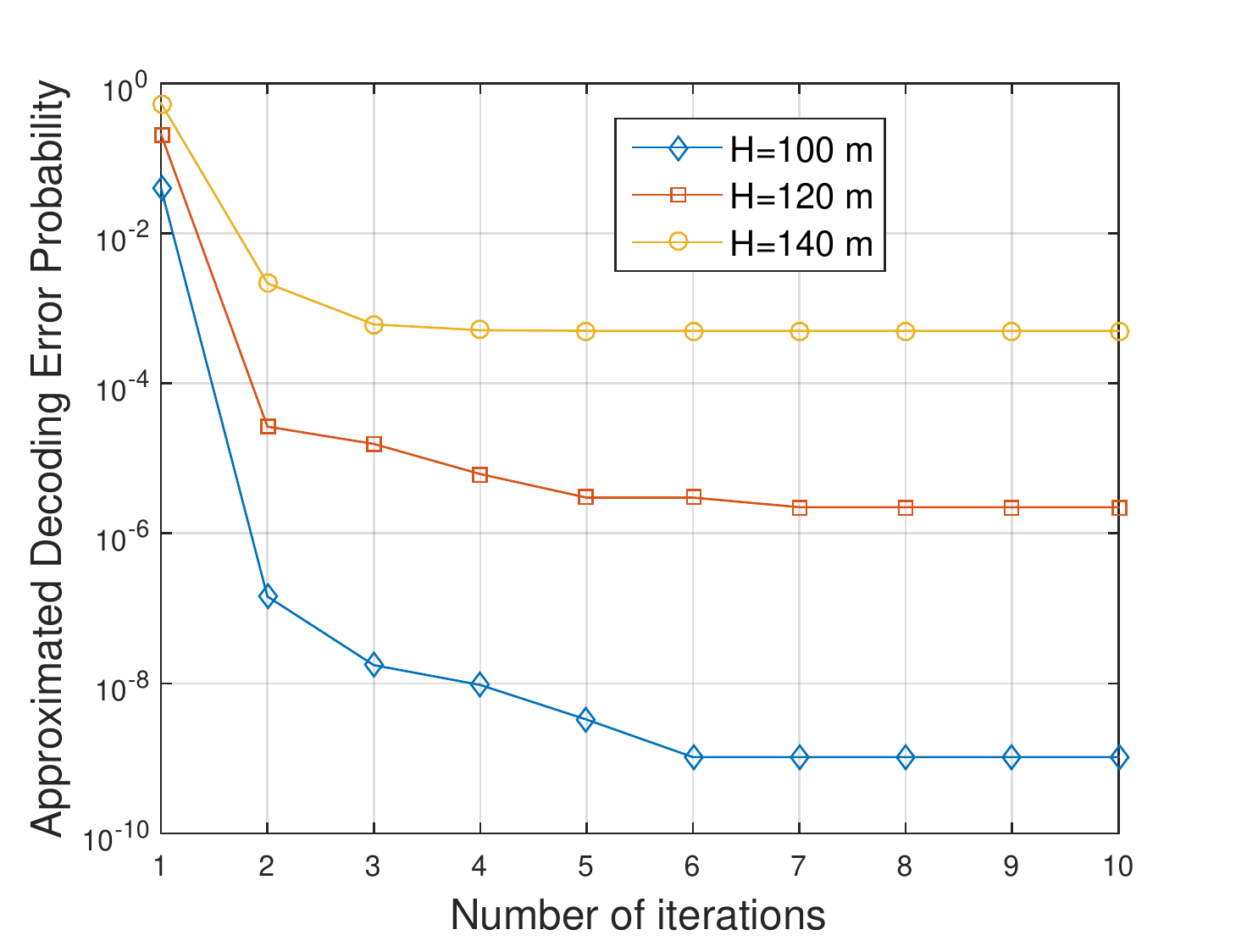}
\vspace{-0.2cm}
\caption{Convergence behaviour of Algorithm \ref{caffrgf} for various $H$. }
\vspace{-0.4cm}
\label{afhishgugt}
\end{figure}

In Fig. 3, we plot  the convergence behaviour of Algorithm \ref{caffrgf}  for various $H$. It is shown in Fig.~\ref{afhishgugt} that the algorithm converges rapidly and generally ten iterations are enough for convergence for all considered $H$, which indicates that our algorithm has a low complexity.

\begin{figure}
\centering
\includegraphics[width=2.4in]{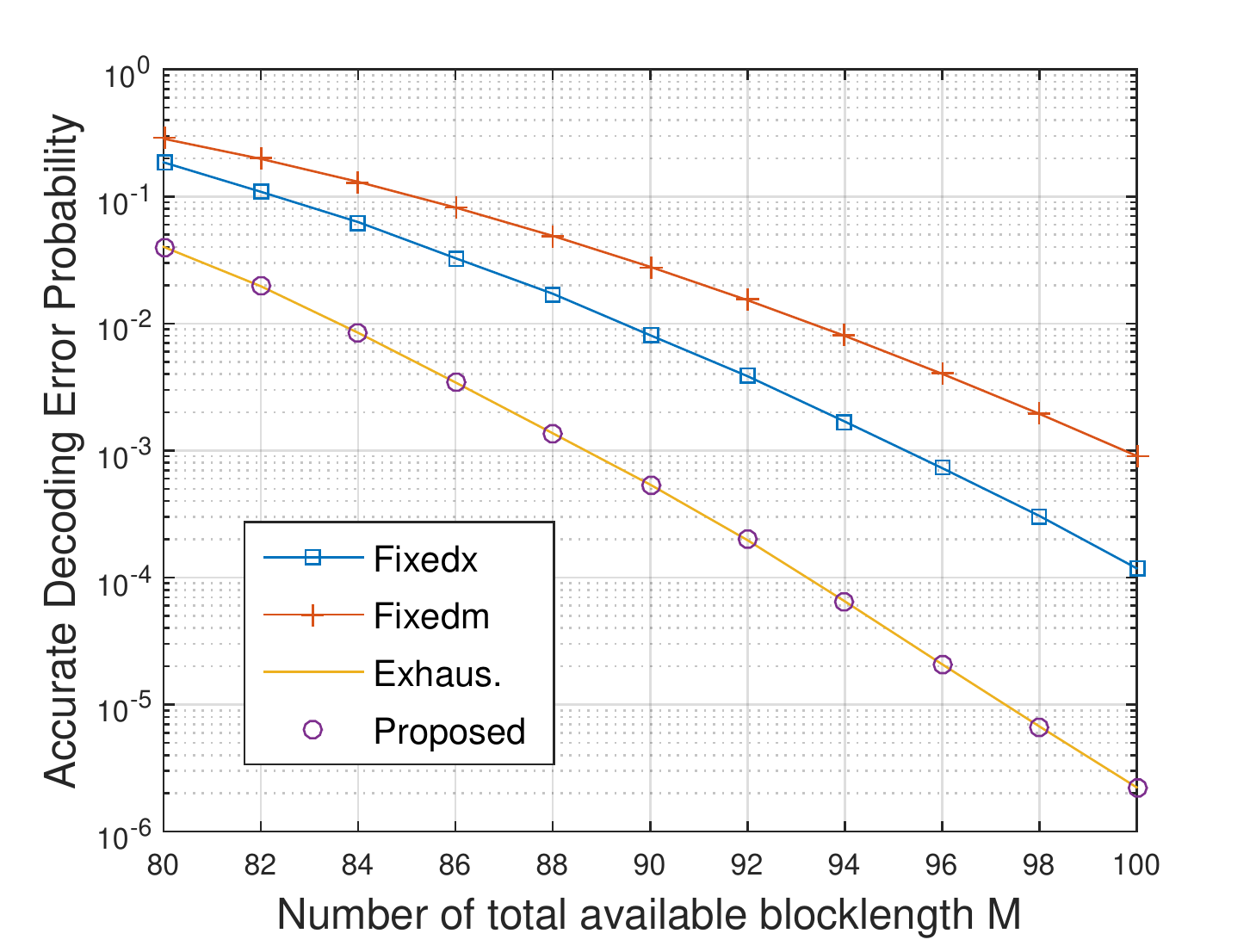}
\vspace{-0.2cm}
\caption{Performance comparison for various algorithms. }
\vspace{-0.4cm}
\label{performancecom}
\end{figure}

In Fig.~\ref{performancecom}, we compared  the proposed algorithm with the following algorithms: 1) exhaustive search algorithm (labeled as `Exhaus.'), 2) optimal blocklength allocation with fixed location $x=(d_1+d_2)/2$ (labeled as `Fixedx'), and 3)  the optimal location with fixed blocklength $m_1=M/2$ (labeled as `Fixedm'). It is seen from Fig.~\ref{performancecom} that the proposed algorithm achieves  the same performance as that of the  exhaustive search algorithm, and significantly outperforms the other two algorithms, which emphasizes the importance of joint optimization.

\vspace{-0.2cm}
\section{Conclusions}\label{conclu}

This paper studied the joint location and blocklength allocation for UAV relay system with URLLC requirement. An effective low-complexity iterative algorithm was proposed to solve the optimization problem. Each subproblem can be solved by using the bisection search method. Simulation results confirm that the proposed algorithm achieved the same performance as that of the exhaustive search method, and has superior performance over the existing algorithms.


\
\






\vspace{-0.2cm}
\bibliographystyle{IEEEtran}
\vspace{-0.2cm}
\bibliography{myre}


\end{document}